\documentclass{jfm}
\usepackage{float}
\usepackage{graphicx}% Include figure files
\usepackage{dcolumn}% Align table columns on decimal point
\usepackage{bm}% bold math
\usepackage{natbib}
\usepackage{epstopdf, epsfig}
\usepackage{array}
\usepackage{tabularx}
\usepackage{enumerate}
\usepackage{amsmath}
\usepackage{slashed}
\usepackage{wrapfig}
\usepackage[dvipsnames]{xcolor}
\usepackage{comment}
\usepackage{multirow}

\shorttitle{Front propagation in a compliant channel}
\shortauthor{C. Cuttle, D. Pihler-Puzovi\'c and A. Juel}

\title{Dynamics of front propagation in a compliant channel}

\author{Callum Cuttle, Draga Pihler-Puzovi\'c
\and Anne Juel \corresp{\email{anne.juel@manchester.ac.uk}},}

\affiliation{MCND and School of Physics \& Astronomy, University of Manchester, Oxford Road, Manchester M13 9PL, UK}

\begin{document}
\maketitle

\begin{abstract}
Front-propagating systems provide some of the most fundamental physical examples of interfacial instability and pattern formation. However, their nonlinear dynamics are rarely addressed. Here, we present an experimental study of air displacing a viscous fluid within a collapsed, compliant channel -- a model system for pulmonary airway reopening. We show that compliance induces fingering instabilities absent in the rigid channel and we present the first experimental observations of the counter-intuitive `pushing' behaviour previously predicted numerically, for which a reduction in air pressure results in faster flow. We find that pushing modes are unstable and moreover, that the dynamics of the air-fluid front involves a host of transient finger shapes over a significant range of experimental parameters.
\end{abstract}


\section{Introduction}

Morphological growth ranging from tumour angiogenesis~\citep{Giverso2016} and growth~\citep{Bru2003}, to bacterial colonies~\citep{Golding1998} and electrodeposition \citep{Schneider2017} is susceptible to interfacial instabilities which can lead to pattern formation and the emergence of disordered dynamics~\citep{Couder2000}. A canonical example is the viscous fingering instability which occurs when air displaces a viscous fluid in the narrow gap between two rigid parallel plates~\citep{saffman1958,paterson1981}. Its complex behaviour stems from the nonlinearity of the interface shape, but has yet to be addressed in terms of its nonlinear dynamics. 

\begin{figure}
\center{\includegraphics[width=0.5\linewidth]
{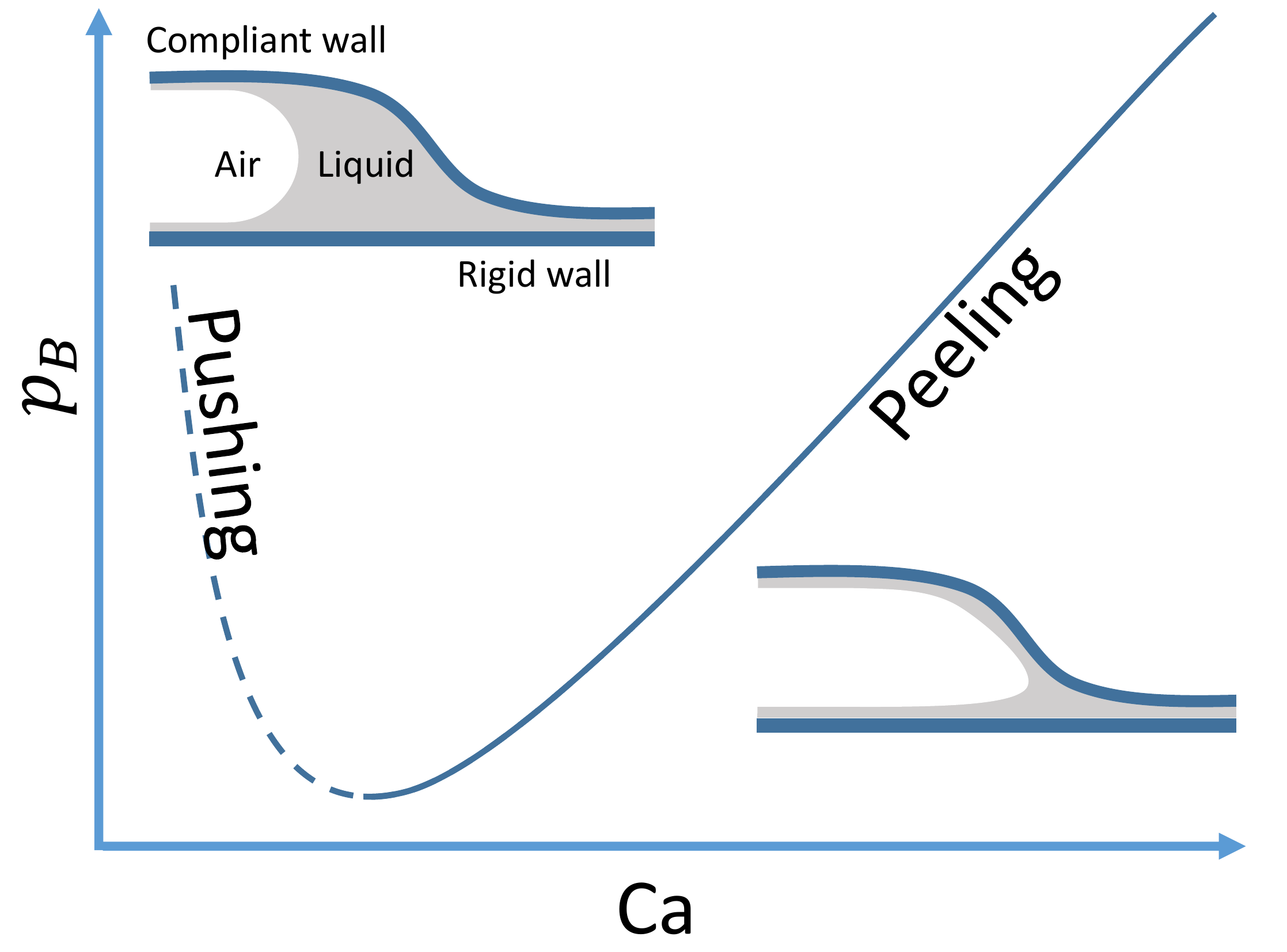}}
\caption{\label{fig:push_peel_schem}Schematic representation of generic push/peel solution branches computed in airway reopening problems in terms of bubble pressure $p_B$ and capillary number $Ca$. The inset figures show longitudinal cross sections, illustrating the typical physical features of pushing and peeling solutions; for pushing modes, the air-liquid interface propagates behind a large volume of liquid, while for peeling modes the interface advances into the most collapsed region of the airway. We have chosen to depict a vessel with a rigid base, of the type studied in this paper.}
\end{figure}

In radial geometries, the growth of the interface is disordered with continually-evolving branched structures arising from initially axisymmetric fronts~\citep{paterson1981}. In contrast, when the flow is confined to a rectangular channel, the system supports a single stable mode of displacement, with planar fronts evolving to a steadily propagating finger of air \citep{saffman1958}. Recent work has shown that in a radial cell the instability may be suppressed or modified if one of the plates is replaced by a compliant membrane, yielding either a stable axisymmetric front or numerous stubby fingers along the interface, depending on flow rate~\citep{PihlerPuzovi2012,Juel2018}. In this paper, we present an experimental study of displacement flows in a rigid rectangular channel with an elastic upper boundary; we show that in this geometry, compliance gives rise to multiple new modes of propagation.

Related systems have been explored in the context of pulmonary airway reopening~\citep{Gaver1990,Heap2008,Duclou2017a,Grotberg2004,Heil2011}. Here, bronchioles, initially collapsed and filled with viscous fluid, were modelled as compliant channels or tubes, reopened by the propagation of an injected finger of air which redistributes the fluid. Numerical studies conducted at fixed levels of initial collapse and compliance have consistently reported a generic two-branch push/peel solution in terms of bubble pressure~$p_B$ and capillary number $Ca$, the non-dimensional finger speed~\citep{Gaver1996,Heil2000,Hazel2003}, as shown schematically in figure \ref{fig:push_peel_schem}. For peeling modes, an increase in $p_B$ drives an increase in $Ca$, while for the pushing modes predicted at low $Ca$ the converse is true -- faster reopening occurs at lower pressures. The pushing branch has been found to be linearly unstable~\citep{Horsburgh2000} in a two-dimensional (2D) geometry consisting of a planar, flexible-walled channel. Fixing the volumetric flow rate $Q$ and allowing $p_B$ and $Ca$ to vary, gives rise to transient relaxation oscillations on the pushing branch, corresponding to abrupt transitions between pushing and peeling~\citep{Halpern2005}, which were characterised using a simplified lumped-parameter model. While this system revealed fundamental dynamics of airway reopening, it lacked any in-plane finger shape due to its reduced dimensionality. Similarly, three-dimensional steady simulations have imposed symmetry about the two longitudinal midplanes of the tube~\citep{Hazel2003}, and so have lacked the multiple propagation modes arising from the fluid-structure interaction flow observed in experiments~\citep{Heap2008}. In strongly collapsed vessels, close to or beyond the point of opposite wall contact, novel asymmetric and symmetric modes were associated with two disconnected peeling branches at low and high $Ca$~\citep{Heap2008,Duclou2017a}.

Consistent with previous experimental studies, we observe two stable peeling branches with different symmetries. A key distinction of the present study is that we are able to provide evidence of a pushing mode, conspicuously absent in previous experimental investigations. The pushing mode occurs transiently for a range of $Ca$ where no stable modes are observed, while straddling a band of $p_B$ over which both peeling branches are stable. Within this region of $p_B$-$Ca$ phase space, the transient evolution of the system towards either stable state is mediated by the unstable pushing modes. These dynamics are reminiscent of the edge states which separate the laminar from the turbulent state in the subcritical transition to turbulence in shear flows~\citep{Barkley2016} or control droplet break-up in sub-critical extensional flows~\citep{gallino2018}. We find that the evolution of the interface in our system is directly influenced by unstable states over a significant range of experimental parameters.

\section{Experimental methods}

\begin{figure}
\center{\includegraphics[width=0.9\linewidth]
{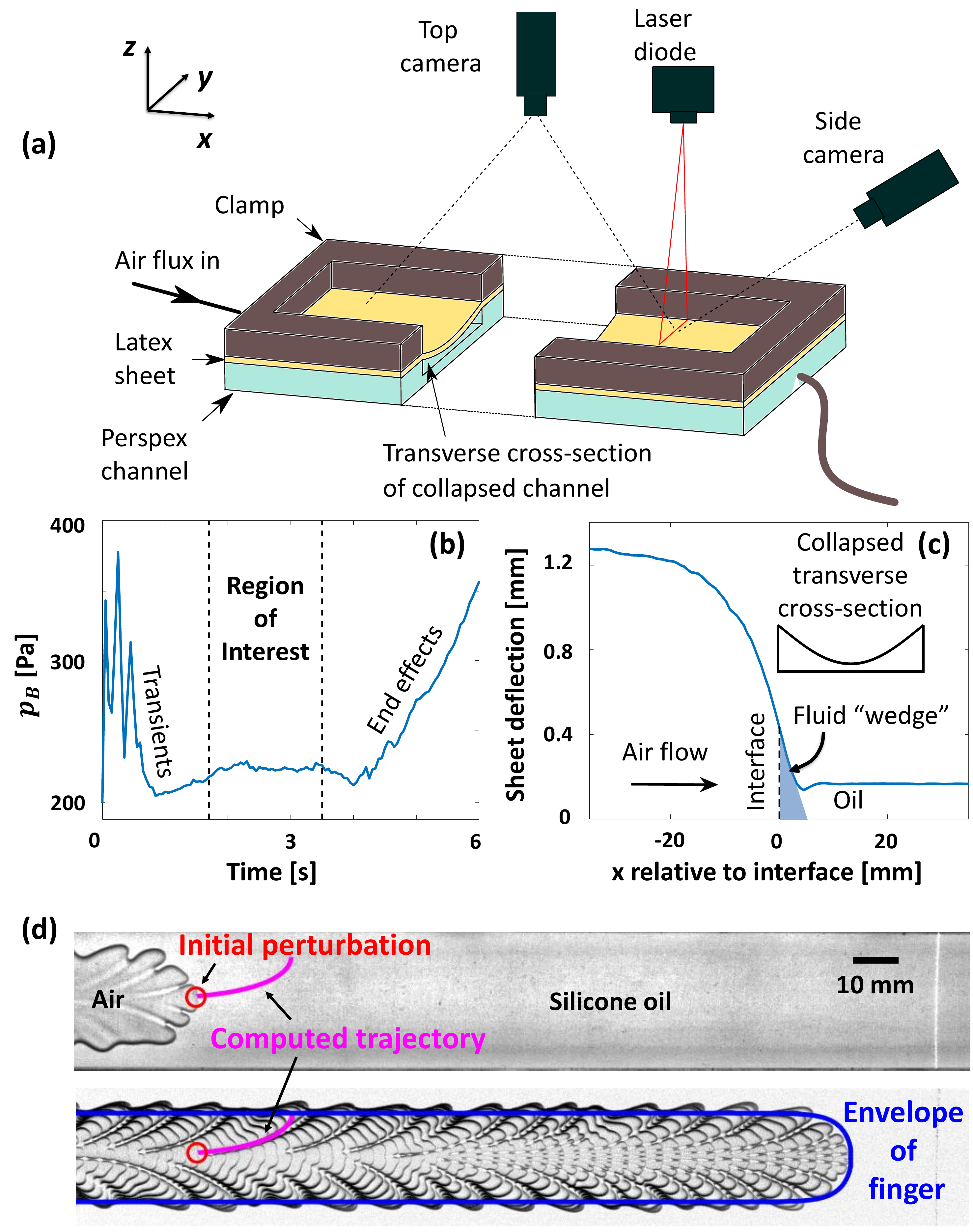}}
\caption{\label{fig:set_up}(a) Schematic diagram of the elasto-rigid channel. (b) Trace of bubble pressure $p_B$ for stable finger propagation, with variations of $<10$~Pa in the region of interest. (c) Instantaneous reopening profile along the midline of the channel walls, centred on the position of the air-oil interface. The inset figure shows schematically a collapsed transverse cross-section. (d) Top to bottom: top-view image of an air finger with short perturbations along the interface; composite image of the same experiment, demonstrating spatio-temporal pattern formation. The envelope of the finger was approximated by the theory of~\citet{saffman1958}. The path of the circled perturbation was fitted following~\citet{lajeunesse2000}, taking the width of the envelope and the initial coordinates of the perturbation as fitting parameters.}
\end{figure}

The experiments were performed in a Hele-Shaw channel with an elastic upper boundary, shown schematically in figure \ref{fig:set_up}(a). A channel of length 60 cm, width $30.00\pm0.02$~mm and depth $1.05\pm0.01$~mm was milled into a block of Perspex, achieving a roughness of less than 10 $\mu$m along the base. The upper boundary comprised a latex sheet (Supatex) of thickness $0.46\pm0.01$~mm, Young's modulus $E=1.44\pm0.05$ MPa and Poisson's ratio $\nu=0.5$. A pre-stress was imposed uniformly along the sheet by hanging evenly distributed weights ($3.03\pm0.01$~kg) from one long edge before clamping the sheet over the channel.

Prior to each experiment, the channel was inflated with silicone oil (Basildon Chemicals Ltd.) of viscosity $\mu = 0.099$~Pa~s, interfacial tension $\sigma = 21$~mN/m and density $\rho=973$~kg/m$^3$ at laboratory temperature $21\pm1^{\circ}$C. The inlet was closed and the oil was allowed to drain via a tube left open to the atmosphere. The initial level of collapse was controlled via the hydrostatic pressure difference $p_{H\!S}$ between the channel and the outlet. In our experiments the channel was strongly collapsed, with the ratio of the transverse cross-sectional area under the elastic sheet to that of the undeformed channel fixed at $A_i=0.44\pm0.02$, close to opposite wall contact which occurs at $A_i=0.36$~\citep{Duclou2017a}. We injected air into the channel at constant volumetric flow rate $10<Q<180$~ml/min, which led to the propagation of an air finger inflating the channel. Flow was controlled via a syringe pump (KDS 200) fitted with Gastight syringes (Hamilton), with initial conditions set by a small precursor bubble of air (less than 1 ml) injected immediately before each experiment. The bubble pressure $p_B$ in the air finger was measured with a pressure sensor (Honeywell $\pm$~5"~H$_2$O), with a typical trace shown in figure \ref{fig:set_up}(b). The experimental region of interest (ROI) covered a 250~mm long region over which pressure variations of less than 10~Pa were recorded during steady finger propagation. This level of accuracy relied sensitively on the uniformity of the pre-stress imposed on the membrane. The air finger was imaged from above by a camera with a spatial resolution of $4.8\pm0.1$~pix/mm, recording 4--30 frames per second (fps), depending on $Q$. The channel was lit from below. We recorded the position of the tip of the air finger in each frame using image analysis routines (MATLAB 2016a), and calculated the tip propagation speed $U$ between frames. Over the ROI, $U$ typically varied by 5\% for steady reopening [top image in figure \ref{fig:peeling_composites}(a)]. 
%\textbf{Which state are you referring to here, it seems like a lot?.} 
The deflection of the membrane was measured using a laser sheet projected across the width of the channel at the end of the ROI. This line was imaged at an oblique angle by a second camera ($22.9\pm0.2$ pixels/mm; 40-160~fps). Membrane deflection at the midpoint of the channel, along with knowledge of the tip propagation speed were then used to reconstruct the profile of the sheet during reopening, as in figure \ref{fig:set_up} (c). Behind the air-oil interface the channel inflates while far ahead the channel is still collapsed; hence, the interface advances into a tapered geometry, behind a wedge-shaped volume of fluid.

The size of this wedge decreases as $Ca$ increases. At high $Ca$, the interface therefore lies within the strongly tapered region, with a printers'-like instability generating multiple viscous fingering perturbations along the tip~\citep[see \textit{e.g.}][]{Duclou2017b,McEwan1966}, as in figure \ref{fig:set_up}(d). Due to the curvature of the interface, these perturbations are advected away from the tip along trajectories instantaneously normal to the interface as it propagates, consistent with the kinematic boundary condition~\citep{lajeunesse2000}. Repeated shedding of perturbations produces complex spatio-temporal patterns, which we visualise using composite images generated from successive frames of a top-view recording, as shown in figure \ref{fig:set_up}(d). Here, the pink trajectory obtained following~\citet{lajeunesse2000} provides a close approximation to the path of the circled perturbation. The separation between advected perturbation trajectories reflects the rate at which perturbations were produced at the tip. Observations of pattern formation provided a sensitive measure of bubble pressure and capillary number, defined here as $Ca=\mu U/\sigma$. Small experimental variations in these parameters were clearly reflected in the pattern, and distinct patterns were associated with each mode of propagation.

\section{Results}

\begin{figure}
\center{\includegraphics[width=0.75\linewidth]
{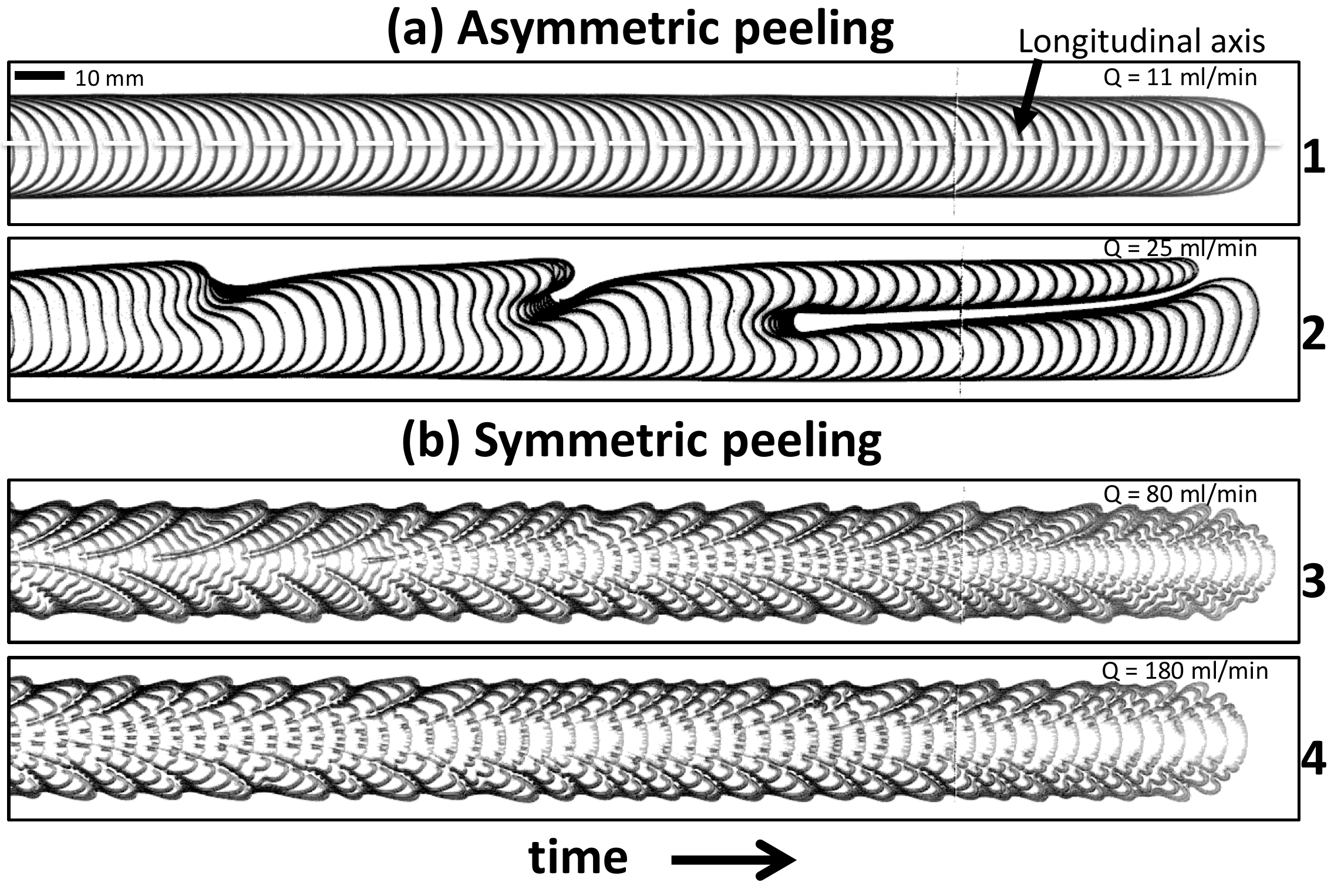}}
\caption{\label{fig:peeling_composites} Composite images of (a) asymmetric and (b) symmetric peeling modes for different values of $Q$, showing the evolution of the interface. The dashed line indicates the axis of symmetry. The asymmetric mode at $Q=11$ ml/min propagated steadily with the tip offset from this axis by a small but significant distance of $6\pm1$\% of the channel's width. Images were recorded at fixed time intervals of (from top): 0.50, 0.25, 0.07, 0.05~s. Numbering is consistent between all figures.}
\end{figure}

\begin{figure}
\center{\includegraphics[width=1\linewidth]
{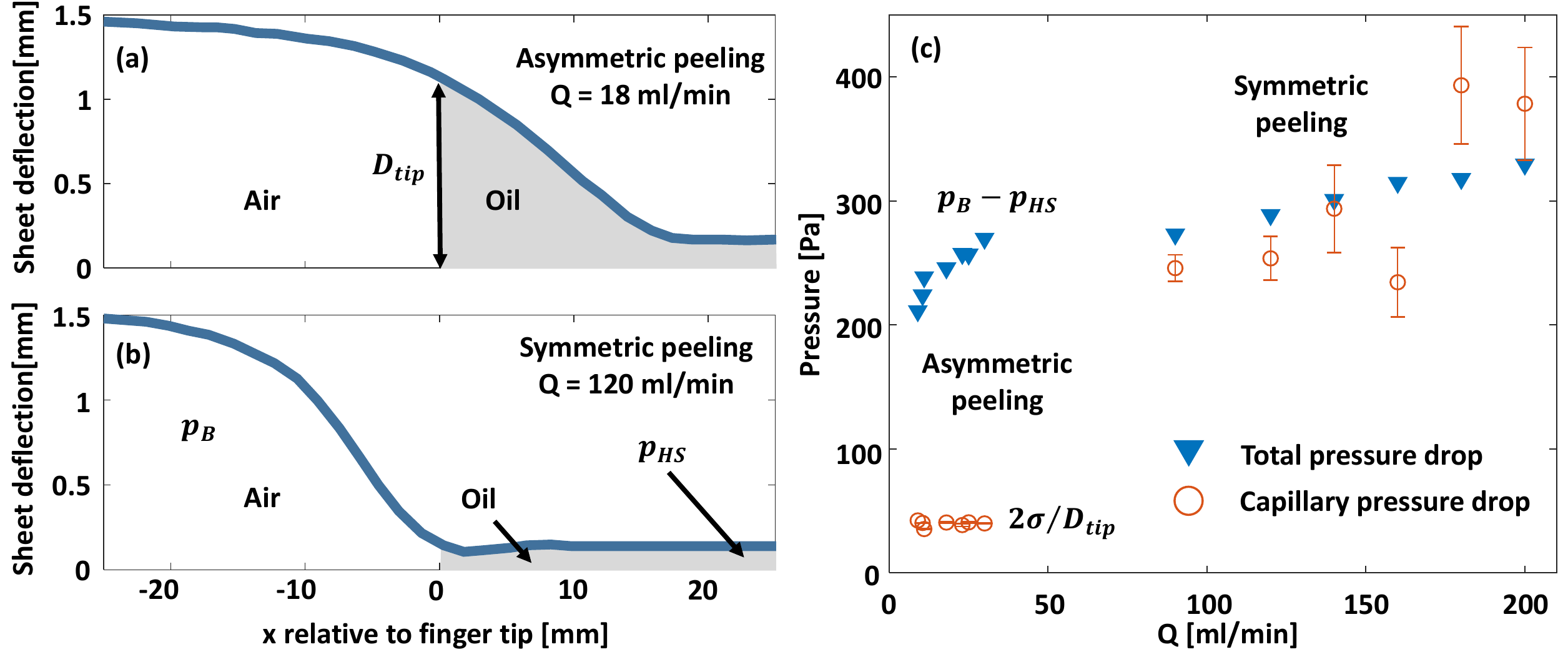}}
\caption{\label{fig:sheet_profiles}(a)-(b) Instantaneous sheet deflection centred on the air-oil interface for (a) asymmetric and (b) symmetric peeling modes. The flow rate $Q$ is given in each figure. The bubble pressure $p_B$ is assumed uniform, while the hydrostatic pressure $p_{H\!S}$ in the oil far ahead of the interface sets the initial collapse of the channel. $D_{tip}$ is the sheet deflection at the interface. (c) Comparison between viscous and capillary effects for each peeling mode. The total pressure drop is $p_B-p_{H\!S}$. The capillary pressure drop is taken to be $2\sigma/D_{tip}$.}
\end{figure}

In our experiments we observed two peeling modes of finger propagation illustrated in figure~\ref{fig:peeling_composites} -- asymmetric or symmetric about the longitudinal axis of the channel. The taper-induced instability described above is unique to the symmetric peeling modes at high $Ca$ [figure~\ref{fig:peeling_composites}(b)]. The small-amplitude fingering perturbation which generates this feathered pattern is characteristic of displacement flows in compliant geometries~\citep{Juel2018}. By contrast, the large indents on the interface in figure~\ref{fig:peeling_composites}(a) due to tip-splitting indicate insensitivity to the tapered geometry at low $Ca$. The distinction lies in the size of the fluid wedge ahead of the interface, which is significantly larger at low $Ca$, as shown in figure~\ref{fig:sheet_profiles}(a,b). As a result, the effects of the compliance-induced taper are diminished and the flow is dominated by viscous dissipation within the wedge [see figure~\ref{fig:sheet_profiles}(c)]. The asymmetric fingers at low $Ca$ arise because of the central constriction of the channel ahead of the interface, which is associated with the quartic profile of the collapsed membrane in the transverse cross-section~\citep{Duclou2017a}; asymmetric fingers minimise viscous dissipation in rigid channels of similar geometry~\citep{FrancoGmez2016}. The contrast between peeling modes is further illustrated in figure~\ref{fig:sheet_profiles}(c); here the total pressure drop within the oil is given by the difference between the bubble pressure $p_{B}$ and the hydrostatic pressure head $p_{H\!S}$ far ahead of the interface, which sets the initial collapse. The contribution to this pressure drop due to capillarity is taken to be the Laplace pressure $\sigma\kappa$ across the interface with the curvature approximated by $\kappa\approx2/D_{tip}$, $D_{tip}$ being the sheet deflection at the interface. For asymmetric peeling modes the capillary contribution is relatively small and so a large viscous pressure drop must be present within the fluid wedge. For symmetric peeling, however, the two terms are of equal magnitude. Hence, this mode of propagation is largely controlled by elasto-capillary effects. Symmetric and asymmetric peeling modes therefore correspond to elastic- and viscous-dominated regimes of reopening, respectively~\citep{Duclou2017a}.

\begin{figure}
\center{\includegraphics[width=0.75\linewidth]
{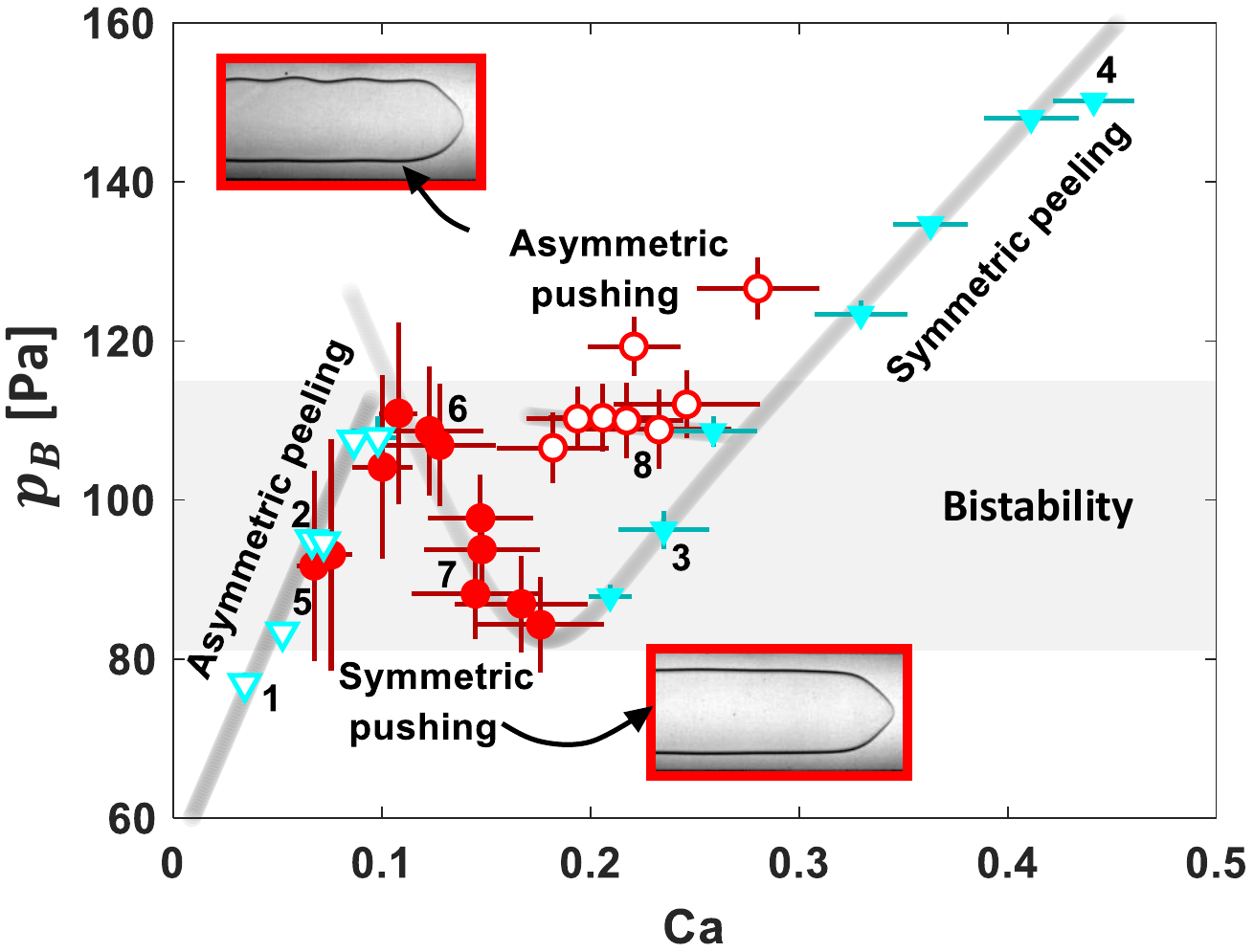}}
\caption{\label{fig:Ca_pressure_curves}Bubble pressure $p_B$ as a function of capillary number $Ca$, showing time-averaged data over the ROI for peeling modes (\textit{triangles}) and transient modes (\textit{circles}). Asymmetric modes are indicated by open symbols, and error bars show the standard deviations of experimental measurements. The interfacial profiles of pushing modes are inset. The band of pressures for which both peeling branches occur is shaded.}
\end{figure}

The pressure of the peeling modes, which is shown with triangles in figure \ref{fig:Ca_pressure_curves} as a function of $Ca$, varies by typically less than the experimental resolution of 10 Pa. This suggests that these modes of propagation are stable in the sense that they persist over the ROI, consistent with peeling modes studied numerically in related geometries~\citep{Halpern2005}. The occurrence of both peeling modes over a similar range of pressures (shaded region) indicates that the system is bi-stable within this band of pressures. 

The two peeling branches are separated by a range of $Ca$ where we do not observe stable finger propagation. Instead, fingers initiated within that range by imposing appropriate values of $Q$ evolve towards either peeling state through the variation of both their pressure and $Ca$. The majority of the time-averaged data of these experiments (\textit{filled circles}) shows a clear trend for bubble pressure to decrease with increasing $Ca$ [figure \ref{fig:Ca_pressure_curves}], which is indicative of a pushing branch [see figure \ref{fig:push_peel_schem}]. However, as we shall discuss below, these data are an average over multiple transient states of the system and do not reflect a single solution branch. Instead, the trend in the time-averaged data reflects the fact that over this parameter range, an increase in the imposed flow rate $Q$ results in an increase in the initial value of $Ca$, while the initial $p_B$ is reduced. The subsequent dynamics of the system depend on these initial conditions and so the $p_B$-$Ca$ relation is reflected in the time-averaged data.

\begin{figure}
\center{\includegraphics[width=0.9\linewidth]
{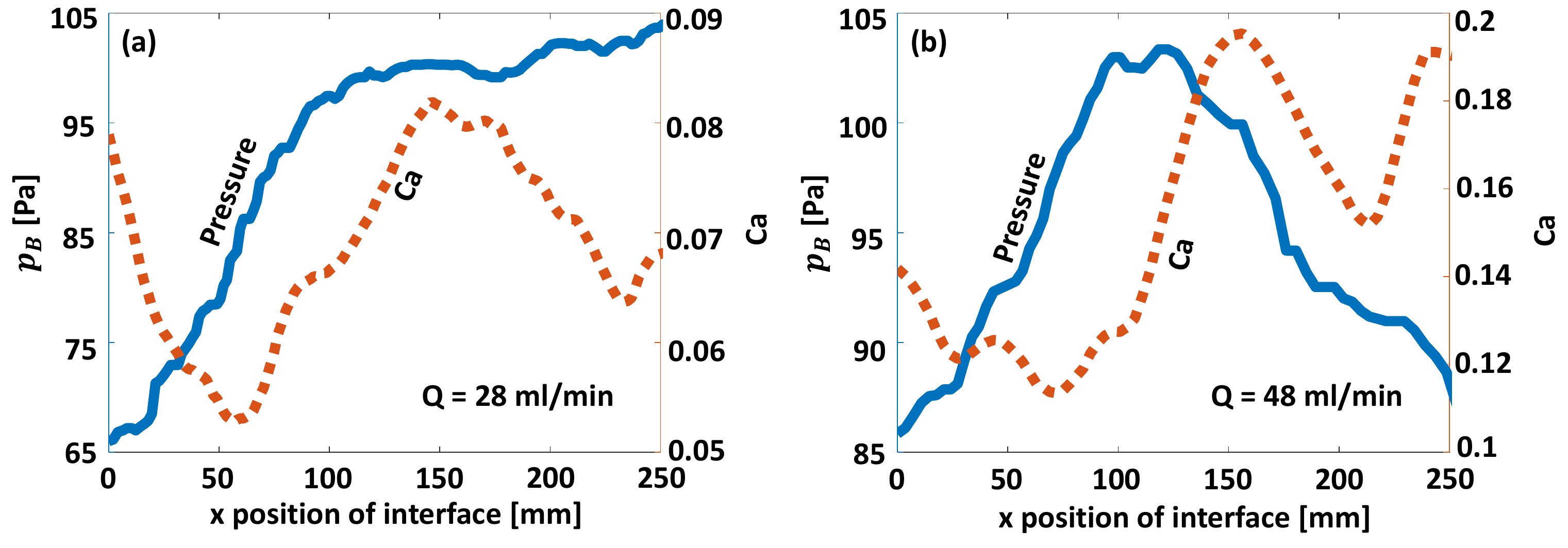}}
\caption{\label{fig:p_ca_vs_x}Capillary number $Ca$ and bubble pressure $p_B$ as a function of the interface position $x$ for two transient modes at fixed flow rate $Q$. (a) and (b) correspond to experiments 5 and 7 in figure \ref{fig:pushing_composites}. The vertical axes in (a) and (b) do not share the same scales.}
\end{figure}

Stronger evidence of the presence of a pushing solution is, however, present upon examination of the system's transient dynamics. The large error bars associated with the transient modes [\textit{circles} in figure \ref{fig:Ca_pressure_curves}] typically reflect a monotonic rise in $p_B$ of 15--40 Pa which occur over at least half of the ROI, as shown for two modes in figure~\ref{fig:p_ca_vs_x}. The rise in $p_B$ is initially accompanied by a decrease in $Ca$, and it is during this phase that we observe symmetric fingers with high (and increasing) in-plane tip curvature, as shown in the lower inset image of figure~\ref{fig:Ca_pressure_curves}. The transient observation of these symmetric pushing modes suggests that they are unstable, as in numerical studies which predict continuous accumulation of fluid in the wedge ahead of the interface~\citep{Halpern2005}. The anomalous group of points at higher $Ca$ (\textit{open circles}) are associated with a similar interface, although here the shape has a definite asymmetry, visible in the upper inset image of figure~\ref{fig:Ca_pressure_curves}. We refer to these as asymmetric pushing modes by analogy.

%We therefore classify these modes as symmetric and asymmetric pushing, accordingly.

%The symmetric peeling mode (\textit{filled triangles}) exhibits variations in the small-amplitude fingering disturbance at the front which result in a slow, approximately periodic variation in $Ca$, with an amplitude indicated by the error bars [figure \ref{fig:peeling_composites}(b)]. In contrast, the asymmetric peeling mode (\textit{open triangles}) propagates at approximately constant $Ca$ even for $Q\ge$ 15 ml/min when tip-splitting occurs, which becomes increasingly irregular as $Q$ increases [figure \ref{fig:peeling_composites}(a)].

%This is to be expected, given that the pushing modes occupy a region of bistability, a band of pressures in which the system may select either of the stable peeling modes, as indicated in figure \ref{fig:Ca_pressure_curves}. 

\begin{figure}
\center{\includegraphics[width=0.75\linewidth,]
{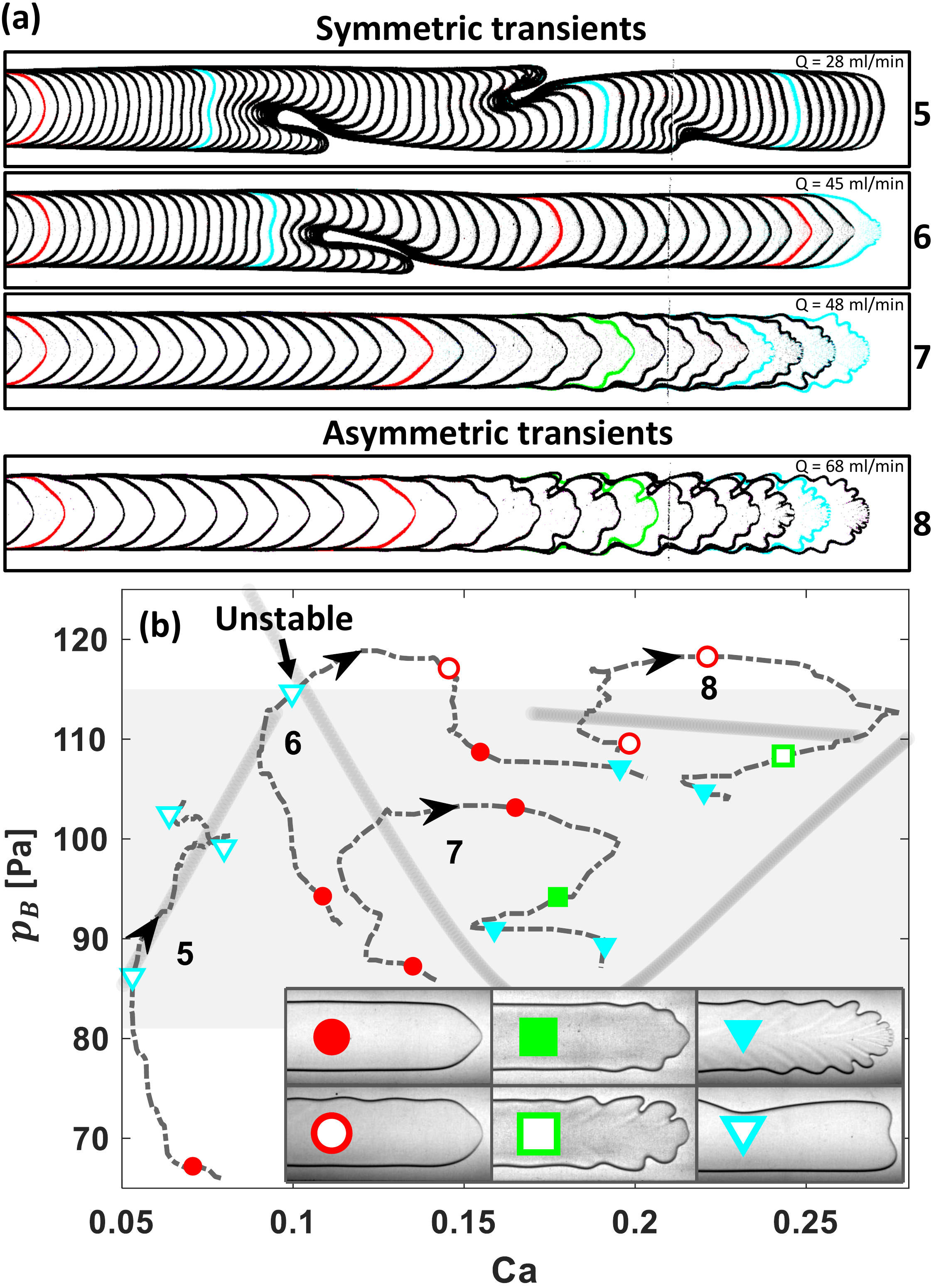}}
\caption{\label{fig:pushing_composites}(a) Composite images showing the transient evolution of fingers towards either the asymmetric or the peeling modes for a range of flow rates $Q$. Images were recorded at fixed time intervals of (from top): 0.28, 0.20, 0.26, 0.26~s. (b) Time-variation of bubble pressure $p_B$ and $Ca$ for each experiment in (a). Arrows indicate increasing time, and symbols correspond to the highlighted interfaces in (a), with pushing (\textit{circles}) and peeling (\textit{squares and triangles}) modes indicated by the key, inset.}
\end{figure}

Figure~\ref{fig:pushing_composites}(a) shows composite images from four representative experiments initiated for values of $Ca$ in the unstable region. While the fingers initially adopt the high curvature profile of the pushing modes identified above, their interfaces evolve as they propagate, transitioning from pushing to peeling modes through distinct evolutions of their interfacial profiles (see highlighted interfaces). The peeling mode selected depends on the flow rate $Q$, which sets the initial conditions of $p_B$ and $Ca$, thus determining the basin of attraction in which the system initially lies. To illustrate this, figure~\ref{fig:pushing_composites}(b) shows the $p_B$-$Ca$ phase space trajectories of the transient modes in figure~\ref{fig:pushing_composites}(a), which were generated by combining the data sets depicted in figure \ref{fig:p_ca_vs_x}. In all cases, the initial behaviour of transient modes is consistent, with the large monotonic rise in $p_B$ accompanied by a deceleration (reduction in $Ca$) as fluid accumulates ahead of the tip, increasing viscous resistance. This $p_B$-$Ca$ relation is associated with pushing behaviour, indicating that the initial evolution of the system is directed parallel to the pushing branch, an observation corroborated by the retention of the pushing interface over this initial transient. We may therefore classify transient dynamics both in terms of their interfaces and their $p_B$-$Ca$ relation at any given time. We will now examine three representative cases, dependent on $Q$.

At low $Q$ (e.g. experiment 5 in figure \ref{fig:pushing_composites}) the system evolves towards the asymmetric peeling branch, and the interface adopts the flat front associated with this mode. The $p_B$-$Ca$ relation is then inverted, as the system begins to climb along the peeling branch and we observe tip-splitting. Over the final third of the ROI, $p_B$ remains steady to within experimental tolerance [see figure \ref{fig:p_ca_vs_x}(a)], and the asymmetric interface is retained, which suggests that the system may have reached a stable state. Further variations in $Ca$ over this stage of the experiment, meanwhile, may be evidence of relaxation oscillations; this is speculatory, however, given the limited length of the ROI.

At flow rates $Q>46$ ml/min (e.g. experiments 7 and 8) the system does not converge onto the asymmetric peeling branch. Instead, the system initially follows orbital trajectories, similar to the relaxation oscillations described by~\citet{Halpern2005} for pushing modes in an analogous 2D system where the in-plane interfacial structure is absent. The mechanisms responsible are likely similar; during the initial transient, the depth of the fluid wedge ahead of the interface increases as fluid is accumulated. This yields a reduction in viscous resistance in this region, and the tip abruptly accelerates as a result. Over this stage, the pushing mode is retained, with the curvature of the tip increasing further, an adaptation which would reduce viscous resistance, allowing fluid in the wedge to escape past the tip. Immediately after, we observe a reduction in both bubble pressure and $Ca$, consistent with peeling [see also figure \ref{fig:p_ca_vs_x}(b) for $x>150$ mm]. Over these transients [\textit{squares} in figure \ref{fig:pushing_composites}(b)], we observe the appearance of undulating structures along the previously flat sections of interface either side of the tip. These interface shapes are distinct from either pushing or stable peeling. Their $p_B$-$Ca$ relation suggests that they may correspond to unstable peeling modes. The dynamics of the system up to this point are therefore dominated exclusively by unstable modes. The final transient is then an abrupt acceleration at fixed pressure, coinciding with a transition to symmetric peeling. These dynamics are robust, qualitatively describing all experiments at $Q>46$ ml/min.

A distinct dynamical behaviour is observed in experiment 6. At $Q=45$ ml/min, the system converges onto the asymmetric peeling branch at around the maximum pressure recorded for stable asymmetric peeling modes. The system therefore fails to stabilise and instead returns to pushing modes, first asymmetric then symmetric, before transitioning to symmetric peeling. The repeated interactions with pushing modes, even after visiting a marginally stable peeling mode, suggest that the pushing modes are weakly unstable, thus possessing a high-dimensional stable manifold. Finally, we note that the behaviour of experiment 6 is only observed over a narrow range of $Q$. At slightly greater or lower $Q$, the system transitions exclusively to either symmetric or asymmetric peeling. Hence, the results of figure \ref{fig:pushing_composites} suggest that the symmetric pushing mode acts as an organising structure, guiding the system towards either of the stable states, reminiscent of an edge state of the system.

%This indicates that at $Q=45$ ml/min, the system is initially close to an edge state, which The 

\section{Conclusion} 

In summary, we have observed two stable, but discontinuous peeling branches, mediated over a region of bistability by an unstable pushing branch. This structure is an unambiguous indication of subcriticality, a key component of the complex nonlinear dynamics associated with turbulent transition in shear flows~\citep{Barkley2016}. Here, in a flow governed by linear equations, where the nonlinearity is entirely contained within the interface (and the elastic membrane), we have been able to study related dynamics experimentally by direct observation of the interface. Furthermore, our measurements provide strong evidence of the dominant role of unstable states in the transient evolution of a front-propagating system. Finally, we note that numerical simulations of a fully-coupled 2D fluid-structure interaction model of the system, currently in development, has so far revealed a rich multiplicity of modes; a direct comparison with experiments is currently under way. The rich behaviour of this conceptually simple front propagating system paves the way for future joint experimental and numerical investigations into subcritical dynamics.

\bibliography{bibliography5}
\bibliographystyle{jfm}

\end{document}